  \documentstyle[prb,aps,multicol,epsfig]{revtex}
\begin{document}
\draft \date{\today} \title
  {Exact solution for the critical state in thin
   superconductor strips with field dependent or anisotropic
   pinning}
\author{Grigorii P.~Mikitik}
  \address{B.~Verkin Institute for Low Temperature Physics \&
   Engineering, National Ukrainian Academy of Sciences,
   Kharkov 310164, Ukraine}
\author{Ernst Helmut Brandt}
  \address{Max-Planck-Institut f\"ur Metallforschung,
  D-70506 Stuttgart, Germany}
\maketitle

\begin{abstract}
 An exact analytical solution is given for the critical state
problem in long thin superconductor strips in a perpendicular
magnetic field, when the critical current density $j_c(B)$
depends on the local induction $B$ according to a simple
three-parameter model. This model describes both isotropic
superconductors with this $j_c(B)$ dependence, but also
superconductors with anisotropic pinning described by a
dependence $j_c(\theta)$ where $\theta$ is the tilt angle of
the flux lines away from the normal to the specimen plane.
\end{abstract}
    \begin{multicols}{2}   
    \narrowtext

\section{Introduction}

  The critical state model \cite{1} for the magnetic behavior
of superconductors with flux-line pinning has proven very
useful \cite{2} though it originally was applied to the simple
(demagnetization-free) {\it longitudinal} geometry of long
superconductors in parallel magnetic field. It took over 30 years
until an analytical solution of the critical state model was
obtained for the more realistic {\it transverse} geometry of thin
superconductors. The solutions were derived for thin
disks \cite{3} and strips \cite{4} in a perpendicular
magnetic field, extending an earlier work on superconductor strips
with transport current,\cite{5} and finally for elliptic-shaped
platelets.\cite{6} Recent detailed numerical work for
strips \cite{7} and disks \cite{8} of finite thickness shows how
the transition from longitudinal to transverse geometry occurs
with changing aspect ratio of the specimen.

  So far, in the transverse geometry all analytical solutions of
the critical state model were
restricted to the Bean model of constant critical current density
$j_c =$ const, but in many experiments $j_c = j_c(B)$ depends on
the local magnetic induction $B$. For example, the simple Kim
model \cite{9} $j_c(B) = j_c(0) / (1 +|B|/B_0)$ was considered
in many experimental and theoretical papers, see e.g.\ the
reviews \onlinecite{2,10,11} and the partly analytical
calculations for thin strips\cite{12} and disks\cite{13}. While
numerical computations easily allow us to consider any $j_c(B)$
dependence,\cite{7,8,11,14} an exact analytical solution of
some model may give deeper insight since it yields explicit
dependences of the resulting quantities on the input parameters.

  In the highly anisotropic
high-$T_c$ superconductors the flux-line pinning in general depends
on the angle $\theta$ between the local direction of the magnetic
induction ${\bf B}$ and the $c$ axis, which in typical experiments
is normal to the plane of the sample. For example, this
type of anisotropy occurs when one takes into account the
intrinsic pinning exerted by the CuO planes or the pinning
by extended defects. \cite{11} It has been shown
recently \cite{15,16,17,18} that for thin superconductors of any shape
(with thickness $d$ much smaller than the lateral extension $L$ but
larger than the magnetic penetration depth $\lambda$) any such
out-of-plane-anisotropy of pinning is equivalent to an induction
dependence of the critical sheet current $J_c(B)$ (the sheet
current is defined as the current density integrated over the
film thickness). Thus, the description of the two-dimensional
critical state, e.g., in an anisotropic strip can be reduced
to the analysis of a one-dimensional problem with some $J_c(B)$.
In this case the characteristic scale $B_0$ over which $J_c(B)$
changes is of the order of $\mu_0j_cd$.

  In this paper we present an analytical solution for
the  critical state in thin superconductor strips in perpendicular
field with field dependent critical current density $j_c(B)$ or,
equivalently, with anisotropic pinning described by a $j_c(\theta)$.
A three-parameter model  $j_c(B)$ consisting of two straight lines,
an inclined line at small $B$ and a horizontal line at larger $B$,
is considered. This rather general model is equivalent to a
piecewise constant angular dependence
 $j_c(\theta) =j_{c1}$ for $0        \le \theta < \theta_0$ and
 $j_c(\theta) =j_{c2}$ for $\theta_0 \le \theta < \pi/2$ where
 $j_{c1}$, $j_{c2}$, and $\theta_0$ are the parameters of the
model. We shall show below that the steepness of the flux front
in the superconductor essentially
depends on the anisotropy of pinning. In particular, in the
case corresponding to the intrinsic pinning in high-$T_c$
superconductors, the front is a very sharp step, which should be
taken into account in analyzing data of local magnetic
measurements. We shall also show that under certain conditions
two penetrating flux fronts can occur in an anisotropic
superconductor.

As usual, we consider here the cases when the characteristic
magnetic field in the sample is sufficiently large such that
the difference between the magnetic
induction $B$ and the field $H$ may be disregarded. This condition is
satisfied when $j_c d$ is much larger than
the lower critical field $H_{c1}$ (otherwise, the so-called
geometric barrier \cite{19} must be taken into account).
We shall thus express all the following equations in terms
of the magnetic field $H$,  related to the current density by
the Maxwell equation ${\bf j} = \nabla \times {\bf H}$.

\section{Model and its solution}  

   We consider an infinitely long strip of width $2w$ and thickness
 $d$, filling the space $-w \le x \le w$, $-d/2 \le z \le d/2$,
 i.e.\ we place the $y$ axis of the coordinate system along the central
 line of the strip and the $z$ axis along the external magnetic
 field $H_a$ which is applied normal to the plane of the strip.
 The increasing applied field induces a sheet current $J$ along $y$,
 which is related to the $z$ component of the magnetic field in the
 plane $z=0$ by the Biot-Savart law,
    \begin{equation}  
    H_z(x)=H_a +{1 \over 2\pi}\int_{-w}^w { J(t)\, dt \over t-x} \,.
    \end{equation}
 Here and below all singular integrals are taken in the sense of
 the Cauchy principal value.
 The penetration of the magnetic flux into the
 superconducting strip is described by the following critical state
 equations: In the flux-free central region $|x| \le b(H_a)$ one has
    \begin{equation} 
    H_z=0\,,
    \end{equation}
 while in the region $b(H_a)\le |x|\le w$, where the flux already
 exists, one has
     \begin{equation} 
     |J(x)|=J_c[\,H_z(x)\,] \,.
     \end{equation}
 The position $x=b(H_a)$ of the boundary separating the regions,
 is found by solving these equations. In Eq.~(3) $J_c(H_z)$ is the
 critical value of the sheet current. At present an exact solution
 of Eqs.~(1)--(3) is known \cite{4,20,21} only for the Bean critical
 state model where $J_c=$ const. Below we shall obtain the exact
 solution for the more general case when $J_c(|H_z|)$ has
 the model form (see Fig.~1):
     \begin{eqnarray} 
     J_c(H_z)&=& J_{c1} -  \gamma H_z  ~~~{\rm for}~~
         0\le H_z \le H_z^0 \,, \nonumber \\
     J_c(H_z)&=& J_{c0}    ~~\,~~~~~~~~~~{\rm for}~~~~~~~
         H_z \ge H_z^0 \,.
     \end{eqnarray}
 Here $\gamma=(J_{c1}-J_{c0})/H_z^0$; the
 three parameters $J_{c1}$,  $J_{c0}$, and $H_z^0$ may have any
 positive value.

 As was mentioned above, in the case of thin superconductors
 the dependence of the critical current density $j_c$
 on the angle $\theta$ between the local
 direction of the magnetic induction and the normal to the strip
 plane can be taken into account if one considers this
 superconductor as infinitely thin but with an $H_z$ dependent
 sheet current. The model dependence described by Eqs.~(4)
 corresponds to the following $\theta$-dependence of the critical
 current density \cite{18} shown in Fig.~1:

\begin{figure}[F1]  
\epsfxsize= 0.60\hsize  \vskip 1.5\baselineskip
\centerline{ \epsffile{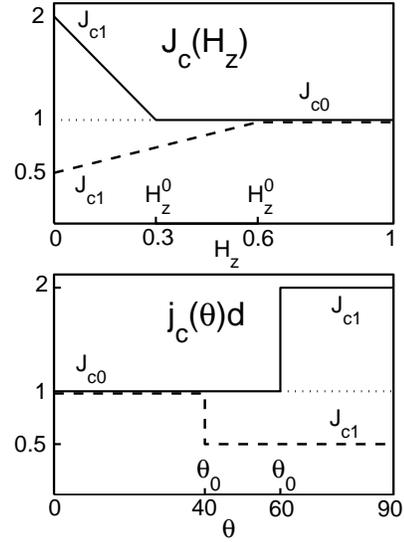}}
                       \vskip 0.5\baselineskip
 \caption{Visualization of the dependence of the critical
sheet current on the perpendicular magnetic field [\,$J_c(H_z)$,
Eq.~(4), upper plot], equivalent to an out-of-plane anisotropy
[\,$j_c(\theta)$, Eq.~(5), lower plot]. The model has three
independent positive parameters, $J_{c0}$, $J_{c1}$, and $H_z^0$,
all of same dimension. In this plot we put $J_{c0}=1$ and show
two examples: $J_{c1}=2$ (intrinsic pining, solid lines) and
$J_{c1}=0.5$ (dashed lines),
with $H_z^0 = 0.3$ (0.6) equivalent to
$\theta_0 =\arctan (J_{c0}/2H_z^0) \approx 60$ (40) degrees.
   } \end{figure}  

     \begin{eqnarray} 
     j_c(\theta)&=& J_{c0}/d  ~~~~{\rm for}~\,~~~~~
         0 \le \theta \le \theta_0 \,, \nonumber \\
     j_c(\theta)&=& J_{c1}/d  ~~~~{\rm for}~~~
         \theta_0\le \theta \le \pi/2 \,,
     \end{eqnarray}
 where $\tan\theta_0=J_{c0}/2H_z^0$. Thus, the case $\gamma>0$
 describes intrinsic pinning by the CuO planes in high-$T_c$
 superconductors\cite{22} ($j_c$ peaks at $\theta=\pi/2$),
 whereas the case $\gamma<0$ can be used to analyze
 pinning by columnar defects normal
 to the film ($j_c$ peaks at $\theta=0$).
 In both these cases one can find {\it two-dimensional} solutions
 of the critical state equations for strips of small but finite
 thickness using the results obtained below and Eqs.~(5,6,9--11)
 of Ref.\ \onlinecite{18}.

   Accounting for the symmetry of the sheet current, $J(-x) = -J(x)$,
 we seek the solution of Eqs.~(1)--(4) in the form
     \begin{equation} 
     J(x) = -{x \over |x|}[J_0(x)+J_1(x)] \,
     \end{equation}
 where
     \begin{eqnarray} 
     J_0(x) =J_{c0}\,, ~~~~~~~\,~~~~~~~~~~~~~~~~~~~~~
     b^2 \le x^2 \le w^2 ,\\
     J_0(x) ={2J_{c0} \over \pi}\arctan\left [{(w^2-b^2)\,x^2
     \over w^2 (b^2 -x^2)}\right ]^{1/2}, ~ x^2 \le b^2 \,,
     \end{eqnarray}
 while $J_1(x)$ is a new unknown function. The parameter $b$ defines
 the position of the flux front, i.e., $x=b$ is the point where
 $H_z$ goes to zero. This parameter depends on $H_a$ and must be
 determined together with $J_1(x)$. Both $J_0(x)$ and $J_1(x)$
 (and the magnetic field below) are even functions, which
 depend only on $x^2$. The function $J_0(x)$ has the form of the
 exact solution \cite{4} to Eqs.~(1)--(3) in the case when
 $J_c=J_{c0}$ and the external magnetic field is equal to
    \[
    H_b=H_{cs}\,{\rm arccosh}(w/b)
    \]
 where $H_{cs}=J_{c0}/\pi$. Using Eqs.~(1), (6)--(8),
 the expression for the magnetic field can be rewritten as
     \begin{equation}  
     H_z(x) =H_0(x) -{1 \over 2\pi}\int_0^{a^2}\!\!
     {J_1(\sqrt s)\,ds\over  s -x^2} \,,
     \end{equation}
 where $a$ is defined by the equality $H_z(a)=H_z^0$, and $H_0(x)$
 is the sum of $H_a$ and the field generated by the current
 $J_0(x)$,\cite{4}
     \begin{eqnarray} 
     &&H_0(x) =H_a-H_b\,,~~~~~~~~~~~~~~~~~~
     0 \le x^2 \le b^2\!,\\
     &&H_0(x) =H_a-H_b + ~~~~~~~~~~~~~~~~~~~~~~~~~~~
     ~~~~~~~~~ \nonumber \\
     && H_{cs} {\rm arctanh}\left [{(x^2 -b^2)\,
     w^2 \over x^2 \, (w^2 -b^2)}\right ]^{1/2} \!\!,~~~
     b^2\le x^2 \le w^2\!\!.
     \end{eqnarray}
 In Eq.~(9) it was taken into account that $J_1(x)$ differs
 from zero only in the region $0\le x^2 \le a^2$
 where $H_z(x) < H_z^0$.

   With the above formulas, the critical state equations take the
 following form: In the interval $0\le x^2\le b^2$ one has
   \begin{equation}  
   H_0(x) ={1 \over 2\pi}\int_0^{a^2}\!\!{J_1(\sqrt s)\,
   ds\over s-x^2}\,,
   \end{equation}
 and in the region $b^2\le x^2\le a^2$  we arrive at
     \begin{equation}  
     H_0(x) -H_z^0 =
     -{J_1(x) \over \gamma}+
     {1 \over 2\pi}\!\int_0^{a^2}\!\!\! {J_1(\sqrt s)\, ds \over
     s -x^2} \,.
     \end{equation}
 In deriving Eq.~(13) we have expressed $H_z(x)$ for
 $b^2\le x^2 \le a^2$ in terms of $J_1(x)$  using the equality
     \begin{equation} 
     H_z(x)=H_z^0-{J_1(x) \over  \gamma}
     \end{equation}
 that follows from formulas (3), (4), (6), (7).
 Eqs.~(12), (13) are linear singular integral
 equations with Cauchy type kernel. The theory of such
 equations is well elaborated,\cite{23} and hence
 we can find $a$, $b$, and $J_1(x)$ for any given $H_a$.

    To do this, we introduce the following notations:
     \begin{eqnarray} \nonumber
     \alpha\equiv {1\over \pi}\arctan{\gamma\over 2} ,~~\,
     \beta\equiv {1 \over 2}-\alpha \,,  \\ \nonumber
     \alpha_+\equiv \alpha ,~~~~~~~~~~~~ \alpha_-\equiv \alpha+1 ,
                                         \\ \nonumber
     F_{\pm}(t)\,\equiv\,(a^2-t^2)^{\alpha_{\pm}}\,|t^2-b^2|^{\beta}
     \end{eqnarray}
 and define the function $f(t)$ by the equalities
     \begin{eqnarray} \nonumber
     f(t)=-2H_0(t) \,,~~~~~~~~~~~~~~~~~~~~~~~  0\le t<b \,, \\
     f(t)=2\sin\pi\alpha\,\cdot\,[H_z^0-
     H_0(t)] \,,~~~~~b < t \le a \,, \nonumber
     \end{eqnarray}
 i.e., $f(t)$ is discontinuous at $t = b$. Then, the solution
 of Eqs.~(12), (13) can be represented as follows: In the interval
 $0\le x^2 \le b^2$ one has
     \begin{equation}  
     J_1(x) = {2 \over \pi}|x|F_{\pm}(x)\int_0^{a}\!\!\!
     {f(t)\, dt \over (t^2 - x^2)F_{\pm}(t)} \,,
     \end{equation}
 while in the interval $b^2\le x^2\le a^2$ we arrive at
     \begin{eqnarray}  
     J_1(x)\!=\!\cos\pi\alpha \! \left [ f(x)\!+\!
     {\gamma \over \pi}|x|F_{\pm}(x)\!\!\!\int_0^{a} \!\!\!\!
     {f(t)\, dt \over (t^2\! -\! x^2) F_{\pm}(t)} \right ]\!,
     \end{eqnarray}
 and $J_1(x)=0$ for $a^2\le x^2\le w^2$.
 Here the integrals are taken in the sense of the Cauchy principal
 value; $F_+$ and $F_-$ refer to positive and negative values of
 $\gamma$, respectively. If $\gamma <0$,
 for the above solution to exist it is necessary that
     \begin{eqnarray}  
     \int_0^{a}\!\!\! {f(t) \over F_-(t)} dt =0 \,,
     \end{eqnarray}
 and
     \begin{eqnarray}  
     \int_0^{a}\! {t^2 f(t) \over F_-(t)} dt =0 \,.
     \end{eqnarray}
 These two equalities enable us to determine $b$ and $a$ when
 $\gamma <0$. If $\gamma >0$, the necessary condition for the
 existence of the solution is
     \begin{equation}  
     \int_0^{a}\!\!\! {f(t) \over F_+(t)} dt =0 \,.
     \end{equation}
 A second relation between $a$ and $b$ in this case is obtained
 from the analysis of the magnetic field near the point
 $x^2 =a^2$. It turns out that
     \begin{equation} 
     H_z(x) -H_z^0 \approx C_{\pm}{\gamma \over 2|\gamma|}
     (4 +\gamma^2)^{1/2}(x^2 -a^2)^{\alpha_{\pm}}\,
     \end{equation}
 if $x^2$ tends to $a^2$ from above, and
     \begin{equation} 
     H_z(x) -H_z^0 \approx C_{\pm}(a^2 -x^2)^{\alpha_{\pm}}\,
     \end{equation}
 if $x^2$ approaches $a^2$ from below. Here $C_{\pm}$ are certain
 integrals independent of $x$; the subscripts $+$ and $-$
 refer to the cases of positive and negative $\gamma$,
 respectively. Since $H_z(x) \ge H_z^0$  when
 $x^2 >a^2$, we find that $C_{+} \ge 0$. On the
 other hand, one has $H_z(x) \le H_z^0$ when $x^2 < a^2$, and
 thus $C_{\pm} \le 0$.
 Hence, one concludes that $C_{+}=0$.
 This is the second equality in the case of positive $\gamma$,
 and it has the form
     \begin{eqnarray}  
    \int_0^{b}\!\!\! { f(t)\, dt \over (a^2\! -\!t^2)F_+(t)}-
    {f(a)\over 2\alpha a (a^2\! -\!b^2)^{1/2}} + ~~~~~~~~
    \nonumber \\
    \int_{b}^{a}\!\! \left[ {f(t) \over t\,
    (t^2\! -\!b^2)^{\beta}} -{f(a) \over a(a^2\! -\!b^2)^{\beta}}
    \right] {t\,dt \over (a^2\! -\!t^2)^{1+\alpha}} =0 \,.
    \end{eqnarray}
 The Eqs.~(19) and (22) determine $a$ and $b$ when $\gamma>0$.

 \begin{figure}[F2] 
\epsfxsize= 0.95\hsize  
\centerline{ \epsffile{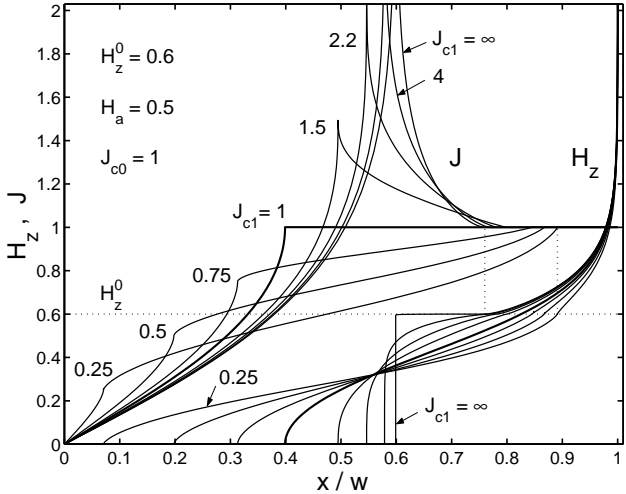}}
                       \vskip 0.5\baselineskip
 \caption{Some profiles of the sheet current $J(x)$ and of the
perpendicular magnetic field $H_z(x)$ in a superconductor thin
strip with width $2w$ for various $J_c(H_z)$ dependences, Eq.~(4),
equivalent to various out-of-plane anisotropies, Eq.~(5), in an
applied field $H_a=0.5$. The unit for both $J$ and $H$  is
$J_{c0}=1$. The anisotropy parameters are $H_z^0=0.6$ and
$J_{c1}=$ 0.25, 0.5, 0.75, 1, 1.5, 2.2, 4, and $\infty$.
The isotropic (or Bean) case $J_{c1}=1$ is shown as bold lines.
The dotted lines indicate the field $H_z = H_z^0$ and the position
$x=a$, where $J(a)=J_{c0}$ and $H_z(a)=H_z^0$. In the limit
$J_{c1} \to \infty$ the field $H_z(x)$ at the flux front $x=b$
abruptly jumps to the value $H_z^0$ and stays constant
for $b \le x \le a$.
   } \end{figure}  

\section{Analysis}

    Let us now analyze the obtained solution. For
 evaluation of the integrals in Eqs.~(9), (15)-(19), (22) we use the
 method given in Appendix A. Some profiles $J(x)$, Eq.~(6), and
 $H_z(x)$, Eq.~(9), obtained in this way are shown in Figs.~2 to 5.

    It should be noted that no restriction on $C_-$ is obtained
 when $\gamma < 0$. In this situation
 the constant $C_-$ is not equal to zero but negative, and thus
 the derivative of $H_z$ with respect to $x$ becomes infinite at
 $x=a$. In the same point a sharp bend occurs in $J(x)$. In other
 words, we obtain that {\it two} flux fronts exist in the sample,
 at $x=b$ and at $x=a$, see Figs.~2, 4, and 5.
 Of course, the singularities in $H_z$ and in $J$ at $x=a$ result
 from the sharp bend in our model $J_c(H_z)$ at $H_z =H_z^0$,
 see Eqs.~(4) and Fig.~1.
 However, one may expect that in the case $\gamma < 0$ our
 qualitative conclusion on the existence of the second flux front
 in the sample remains valid if $J_c(H_z)$ is a smooth function
 but its behavior changes abruptly over an interval smaller than
 $H_{cs}$. Such changes indeed may occur if the critical current
 density has sufficiently sharp angular dependence $j_c(\theta)$.

   We shall now describe $H_z(x)$ and $J(x)$ in the vicinity of
 the point $x=b$ in which $H_z=0$.
 According to Eq.~(14), at this point $|J(b)|=J_{c1}$.
 When $x^2\le b^2$, it follows from
 the exact solution that
      \begin{equation} 
      |J(x)|-J_{c1}\approx C_b^{\pm}(b^2-x^2)^{\beta}\,,
      \end{equation}
 while if $x^2\ge b^2$, one has
      \begin{equation} 
      |J(x)|-J_{c1}\approx {\gamma \over (4+\gamma^2)^{1/2}}
      C_b^{\pm}(x^2-b^2)^{\beta}\,.
      \end{equation}
 Here $C_b^+$ and $C_b^-$ are certain integrals which do not
 depend on $x$ and have negative values. Formulas (23) and (24)
 show that in the case $\gamma>0$, $|J(x)|$ has a sharp peak at
 $x=b$, whereas for $\gamma<0$, $J(x)$ is a monotonic function
 and its derivative with respect to $x$ becomes infinite at $x=b$,
 see Figs.~2, 4, and 5. Taking into account the above formulas and
 Eq.~(14), one obtains the distribution of $H_z$ near $x=b$,
      \begin{eqnarray} 
     H_z&=& 0 ~~~~~~~~~~~~~~~~~~~~~~~~
     ~~~~~~~~{\rm for}~~  x \le b \,, \\
     H_z&=& -{C_b^{\pm} \over (4+\gamma^2)^{1/2}}(x^2-b^2)^{\beta}
     ~~~{\rm for}~~   x \ge b \,.
     \end{eqnarray}
 When $\gamma=0$, we arrive at the well-known result \cite{4,20,21}
 $H_z\propto (x^2-b^2)^{1/2}$. However, in the general case, taking
 into account the equality
 $\beta={1 \over 2}-{1 \over \pi}\arctan(\gamma/2)$,
 one may conclude that the greater $\gamma$ is, the sharper is the
 $H_z$ profile, Fig.~2. Interestingly, the dependence
 $(x-b)^{\beta}$ sufficiently well describes $H_z(x)$ even if $x$
 is not too close to $b$, see Fig.~6.

    Consider now the solution in the limit of small positive
 values of $\gamma$. If $\gamma \rightarrow 0$, two
 cases are possible: $H_z$ remains a constant, or it increases
 as $\gamma^{-1}$ (i.e., $J_{c1}-J_{c0}=$ const.). In the first case
 one has $\alpha \approx \gamma/2\pi$, $H_a-H_b\propto \gamma$,
 and the function $f(t)$ tends to zero. Thus, according to
 Eqs.~(15) and (16), $J_1\rightarrow 0$, and the solution goes
 over to the well-known result \cite{4,20,21} for the Bean
 critical state model with $J_c=J_{c0}$. In the second case
 $J_0(x)+J_1(x)$ also tends to the solution corresponding to a
 constant $J_c$ but now $J_c=J_{c1}$.

    In the limiting case $\gamma\rightarrow +\infty$, this parameter
 drops out from Eqs.~(15), (16), (19), (22), and
 $J_1(x)$ depends only on $H_z^0$, $J_{c0}$. In other words, if
 $J_{c1}\gg J_{c0}, H_z^0$ or $J_{c1}>J_{c0}\gg H_z^0$, the solution
 becomes practically independent of $J_{c1}$. The distribution
 of the magnetic field in this case can be understood using
 Eq.~(26). It turns out that $C_b^+\approx-\gamma H_z^0$ for
 $\gamma \gg 1$, and hence
    \begin{equation} 
     H_z(x)\approx H_z^0(x^2-b^2)^{\beta}\,
    \end{equation}
 with $\beta \rightarrow 0$. This means we have an abrupt step
 of height $H_z^0$ at $x=b$ (see Fig.~2).

   It should be emphasized that this limiting case,
 $\gamma \rightarrow +\infty$,
 corresponds to intrinsic pinning in high-$T_c$
 superconductors in which the ratio
 $[j_c(\pi/2)/j_c(0)]=J_{c1}/J_{c0}$ can be sufficiently large
 (see, e.g., Ref.~\onlinecite{24}).
 Thus, our solution of this limit can be used for
 analyzing the critical state in these superconductors. In particular,
 it follows from Eqs.~(19), (22) that the position of the flux
 front, $b/w$, is a function of $H_a/H_{cs}$ and of the
 parameter $H_z^0/H_{cs}$, see Fig.~7. In general this function
 can {\it not} be
   \linebreak

\begin{figure}[F3]  
\epsfxsize= 0.95\hsize  \vskip 1.5\baselineskip
\centerline{ \epsffile{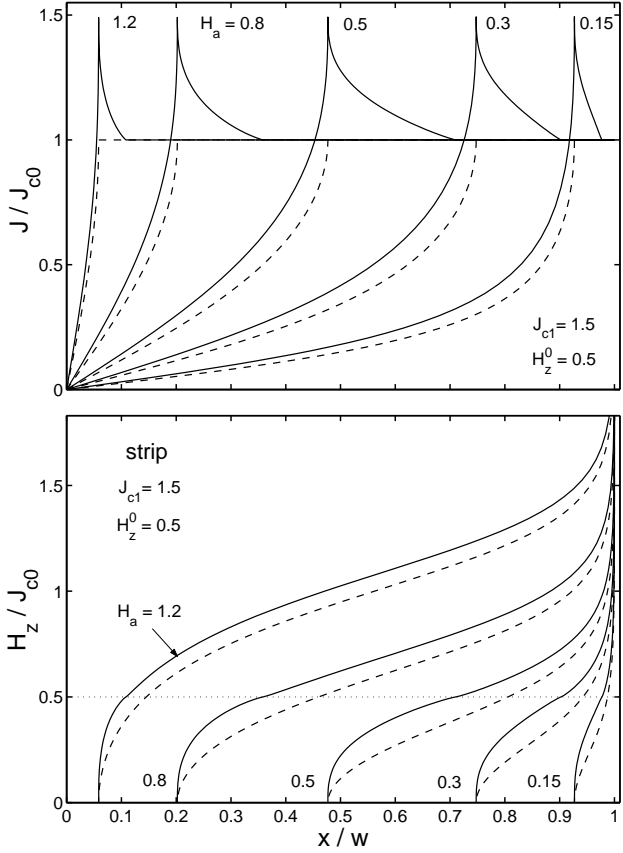}}
                       \vskip 0.5\baselineskip
 \caption{Profiles of the sheet current $J(x)$ (top) and
 of the magnetic field $H_z(x)$ in a thin strip with width
 $2w$ and anisotropic pinning (solid lines) in an increasing
 applied field $H_a=0.15$, 0.3, 0.5, 0.8, and 1.2 in units of
 $J_{c0}=1$. The anisotropy parameters are $J_{c1}/J_{c0}=1.5$
 and $H_z^0/J_{c0} = 0.5$, thus $\gamma =1$. The dashed lines show
 the profiles of an isotropic strip for the same values of the
 front position $b_0(H_a)$, Eq.~(28). Note the sharp peak of
 $J(x)$ at $x=b$ of height $J(b)=J_{c1}$ and the steep front
 of $H_z(x)$ at $x=b$ for this type of anisotropy.
 At $x=a$, $J(x)$ reaches the value $J_{c0}=1$ and $H_z(x)$
 goes through the value $H_z(a)=H_z^0$ marked by a dotted line.
   } \end{figure}  

  \noindent
 fitted by scaling the dependence found in the isotropic case,\cite{4}
   \begin{equation} 
   {b_0\over w}={1\over \cosh(H_a/H_{cs})}\,,
   \end{equation}
 using some effective value of $H_{cs}$. Rather, the shape of
 $b(H_a)$ essentially depends on the ratio $H_z^0/H_{cs}$.
 Therefore,  measuring
 $b(H_a)$ in principle can give information not only on
 $H_{cs}=J_{c0}/\pi$ but also on $H_z^0$, i.e., about the width of
 the peak in $j_c(\theta)$, see Eq.~(5). In particular, when
 $H_z^0\ll J_{c0}$, Eqs.~(19), (22) lead to the following expression
 for the front position:
  \begin{equation} 
   \left({b\over w}\right)^2\!\!\approx {
   1 + k \cdot \left( H_z^0 / H_{cs} \right)^2\!
   \tanh^2(H_a/ H_{cs}) \over \cosh^2(H_a/H_{cs})}\,,
   \end{equation}
 where the constant $k$ is determined by the root of the equation

\begin{figure}[F4] 
\epsfxsize= 0.95\hsize  \vskip 1.5\baselineskip
\centerline{ \epsffile{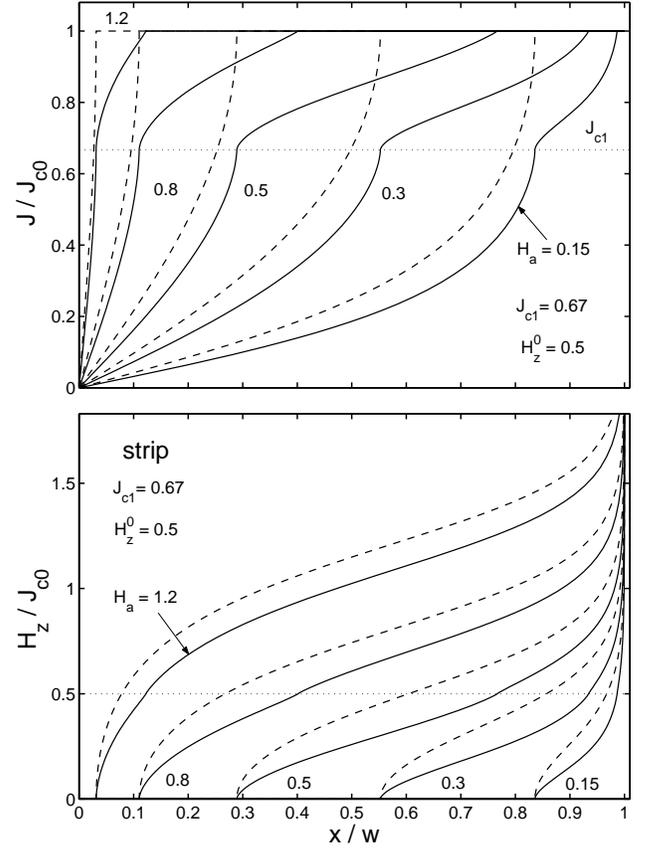}}
                       \vskip 0.5\baselineskip
 \caption{As Fig.~3, but for different type of anisotropy,
 $J_{c1}/J_{c0} =0.67$ and $H_z^0/J_{c0} =0.5$, thus $\gamma=-0.67$.
 In this case $J(x)$ is monotonic and has an inflection point with
 vertical slope at $x=b$ where $J(b)=J_{c1}$ (dotted line). The
 penetrating front of $H_z(x)$ is now less steep than in the
 isotropic case, which is shown as dashed lines.
   } \end{figure}  

    \begin{eqnarray}
    {\pi \over 4}(u^2-1)=u-\arctan u \ , \nonumber \\  \nonumber
    k ={16\over \pi^2}{u^2\over (1+u^2)^2}\approx 0.394 \,.
    \end{eqnarray}
 Note that the right hand side of Eq.~(29) cannot be reduced to
 the dependence (28) in the whole interval of changes of $H_a$
 when $H_z^0$ is different from zero. The exact values of the front
 position $b(H_a)$ are shown in Fig.~7 for the limit of large
 $\gamma \gg 1$, for $J_{c1}=11$ and $H_z^0=0 \dots 1.5$ in
 units of $J_{c0}=1$.

    In the third limiting case when
 $\gamma\rightarrow -\infty$, one has $\beta \rightarrow 1$,
 $C_b^- \sim -H_z^0$ and the the induction profile becomes
    \begin{equation}  
    H_z(x)\propto (x^2-b^2)
    \end{equation}
 with a small prefactor of the order of $H_z^0/|\gamma|$.
 Thus, for $-\gamma \gg 1$ the flux front at $x=b$ practically
 disappears while, according to Eq.~(20), the second front
 near $x=a$ is well developed, see Fig.~5.

\begin{figure}[F5]  
\epsfxsize= 0.95\hsize  \vskip 1.5\baselineskip
\centerline{ \epsffile{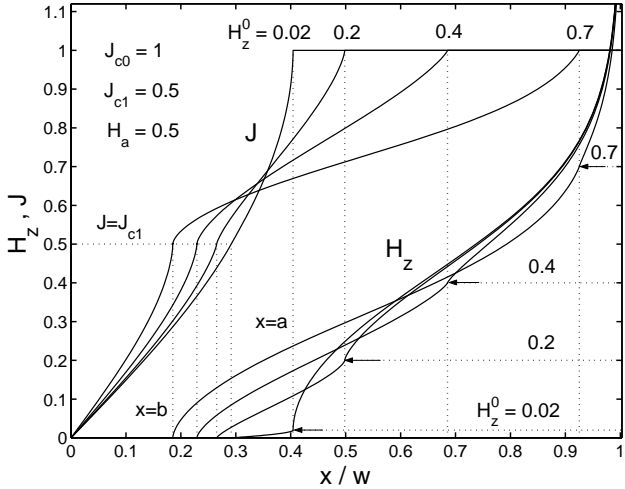}}
                       \vskip 0.5\baselineskip
 \caption{Profiles of $J(x)$ (left) and $H_z(x)$ (right) in a thin
 strip with anisotropic pinning of the type $J_{c1} = 0.5$ for
 various values of $H_z^0 = 0.02$, 0.2, 0.4 and 0.7 in a constant
 applied field $H_a =0.5$ (in units of $J_{c0}=1$). The dotted
 lines at $x=b$, $x=a$, $J=J_{c1}$, and $H_z=H_z^0$ shall help to
 identify the features $J(b)=J_{c1}$,  $J(a)=J_{c0}$,  and
 $H_z(a)=H_z^0$. Note that with decreasing $H_z^0$ the penetrating
 flux front at $x=b$ becomes less pronounced and a new front
 appears at $x=a$. In the limit $H_z^0 \to 0$ only the front
 at $x=a$ remains and the profiles look like in the isotropic strip
 with $b$ replaced by $a$.
   } \end{figure}  

    Finally, we consider in some detail the case of small negative
 values of $\gamma$ when $H_z^0\gg H_{cs}=J_{c0}/\pi $ while
 the ratio $J_{c0}/J_{c1}$ is not close to unity. This case can
 give some idea of pinning by columnar defects, which produce a
 peak in $j_c(\theta)$ at  $\theta=0$. Indeed, if one assumes that
 the characteristic width
 of the peak, $\theta_0$, is small ($\theta_0 \ll 1$), then
 it follows from the definitions of $H_z^0$ and $\gamma$ that
 $H_z^0\approx J_{c0}/2\theta_0$ and $|\gamma| < 2\theta_0$.
 Since the solution with $\gamma =0$ and $J_c=J_{c1}$ describes
 the critical  state in the strip before the irradiation [we assume
 that the columnar defects do not change $j_c(\theta)$ at
 $\theta > \theta_0$], the difference between the solutions
 corresponding to $\gamma\ne 0$ and $\gamma=0$ provides information
 on pinning by columnar defects. In the considered case this
 difference is small, and it can be analyzed analytically.
 In particular, we obtain the following relation between the
 positions of the flux fronts, $b$ and $b_1$, obtained at the same
 $H_a$ in the strip with and without columnar defects, respectively:
   \begin{equation} 
   {\rm arccosh}{w \over b_1} - {\rm arccosh}{w \over b} =
   {|\gamma |\over \pi} \, g(h) \,,
   \end{equation}
 where $h\equiv \pi H_a/J_{c1}$, $w/b_1 =\cosh(h)$, and the function
 $g(h)$ has the form:
    \begin{eqnarray} 
    g(h) = \int_0^h \! \ln (2 \cosh t) \,dt \,.
    \end{eqnarray}
 Since $g$ is a nonlinear function of $h$,
    \begin{eqnarray} 
    g(h) \approx {1 \over 2} h^2 +0.411\, (1-e^{-1.8 h}) \,,
    \end{eqnarray}

\begin{figure}[F6]  
\epsfxsize= 0.95\hsize  \vskip 1.5\baselineskip
\centerline{ \epsffile{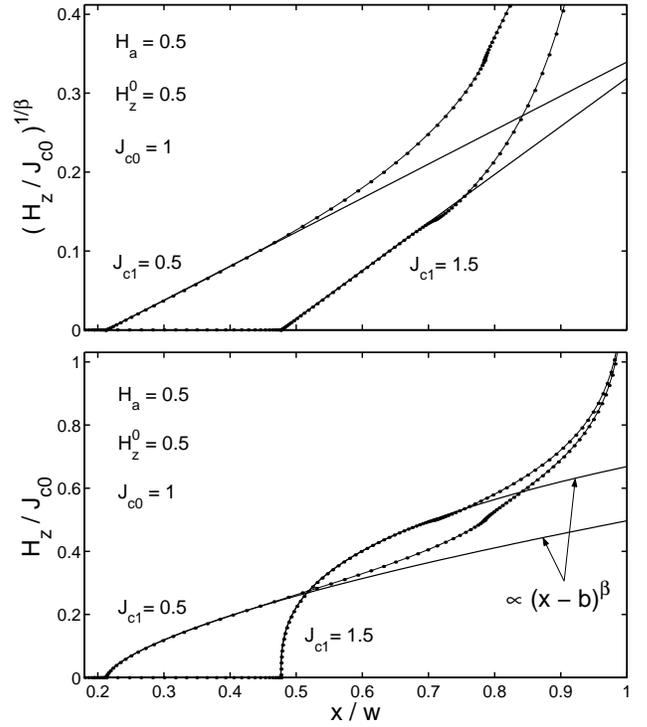}}
                       \vskip 0.5\baselineskip
 \caption{Comparison of the shape of the profile $H_z(x)$ near the
 flux front with the expression $(x-b)^\beta$ suggested by Eq.~(26).
 Shown are examples with $H_a =0.5$ (in units of $J_{c0} =1$) and
 two different anisotropies:  $H_z^0 =0.5$, $J_{c1}=1.5$ (thus
 $\gamma = 1$, $\alpha = 0.148$, $\beta = 0.352$, $b =0.477$,
 $a =0.711$) and  $H_z^0 =0.5$, $J_{c1} =0.5$ (thus $\gamma =-1$,
 $\alpha = -0.148$, $\beta = 0.648$, $b=0.213$, $a=0.784$).
 The exact $H_z(x)$ (dotted lines) is well fitted over a large
 interval of $x$ by the function $c \cdot (x-b)^\beta$ (solid lines)
 with $c=0.840$ or $c=0.580$ for these two examples (with $x$ and $b$
 in units of the strip half width $w$). The solid lines in the
 upper plot are straight lines fitting $H_z(x)^{1/\beta}$.
   } \end{figure}  

 \noindent
 the exact dependence $b(H_a)$ can not be described by Eq.~(28) with
 some effective $H_{cs}$. The prefactor
   \[
   {|\gamma|\over \pi} \approx {2\theta_0\over \pi} \,
   {j_c(0)-j_c(\pi/2)\over j_c(0)}
   \]
 in Eq.~(31) is determined by the characteristics of pinning by the
 columnar defects, i.e., by the width and height of the peak in
 $j_c(\theta)$.

\section{conclusions}

   An exact solution of the critical state equations for
 the strip in perpendicular magnetic field is derived for an
 induction-dependent critical sheet current $J_c(H_z)$
 described by Eqs.~(4). This model dependence may be used to
 simulate the intrinsic pinning by CuO planes
   \linebreak

\begin{figure}[F7]  
\epsfxsize= 0.95\hsize  \vskip 1.5\baselineskip
\centerline{ \epsffile{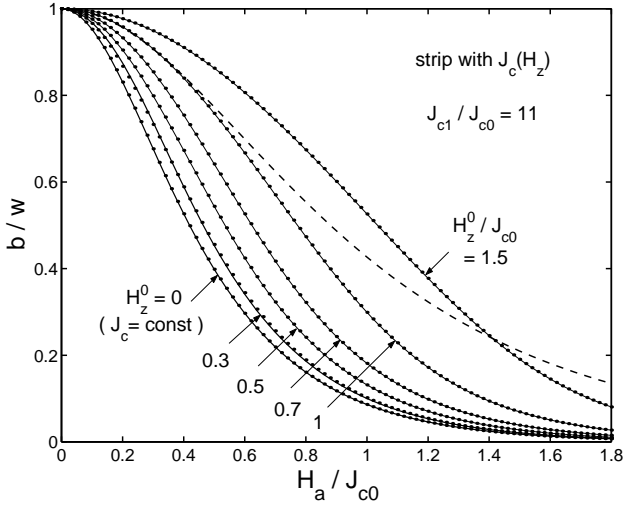}}
                       \vskip 0.5\baselineskip
 \caption{The position $b$ of the flux front, or penetration
depth $w-b$, of a superconductor thin strip with width $2w$
and various $J_c(H_z)$ dependences, Eq.~(4), plotted versus
the applied magnetic field $H_a$ in units of $J_{c0}=1$. The
dotted lines are computed as described in Sct.~III for anisotropy
parameters $J_{c1}=11$ and $H_z^0=$ 0, 0.3, 0.5, 0.7, 1, and 1.5.
The bold solid lines for $H_z^0=$ 0 and 0.3 are from Eq.~(29)
and fit the exact data very well. The dashed line
$b/w = 1/\cosh[H_a/(2.1 H_{cs})]$, obtained
by stretching the isotropic ($J_c=$ const) expression, Eq.~(28),
by a factor of 2.1, demonstrates that such scaling of the
isotropic result cannot fit the anisotropic result.
  } \end{figure}  

   \noindent
 ($\gamma>0$) or pinning by extended defects ($\gamma<0$) in
 high-$T_c$ superconductors. In the case $\gamma>0$, the $H_z$
 profile in the vicinity of the flux front is sharper than in
 the isotropic case, and the current
 density has a sharp peak there. In the limiting case, $\gamma\gg 1$,
 which may describe the intrinsic pinning in high-$T_c$
 superconductors, the field profile $H_z(x)$ has a sharp rectangular
 step. In the opposite situation, $\gamma<0$, {\it two flux fronts}
 can occur in the superconductor; the $H_z$ profile near $x=b$ is
 less steep than in the isotropic case, and the current density is a
 monotonic function of $x$. In both cases of positive and negative
 $\gamma$ the profile $H_z(x)$ in a sufficiently large vicinity of
 the flux front is well approximated by the expression
 $H_z(x)\approx (x-b)^{\beta}$ with the exponent
 $\beta=0.5-\pi^{-1}\arctan(\gamma/2)$.

   The experimental investigation of flux-density profiles near the
 flux front and of the $H_z$ dependence of the penetration depth
 can give information on the strength and anisotropy of flux line
 pinning in superconductors.

\acknowledgments

  G.P.M.~acknowledges the hospitality of the Max-Planck-Institut
f\"ur Metallforschung, Stuttgart.

\appendix{ }

\section{Numerical evaluation}

    The condition that two integrals have to vanish,
 e.g.\ Eqs.~(17,18) of the form $I_1(a,b)=0$ and $I_2(a,b)=0$, we
 satisfy by minimizing the function $U(a,b) = I_1^2 +I_2^2$ with
 respect to $a$ and $b$. After this we calculate the sheet
 current $J_1(x)$ from Eqs.~(15,16) and the magnetic field
 $H_z(x)$ from Eqs.~(9) and (14).

   The integrals (9), (15-19), and (22) over the variable $t$
have integrands which possess one or several infinities at the
points $t=0$, $t=x$, $t=b$ and $t=a$ where the denominators
vanish. We evaluate such integrals in the following way.

    In the integrals containing a factor $(t-x)^{-1}$ we subtract
the singular part and integrate it analytically, e.g.,
   \begin{eqnarray}  
   \int_0^a \!\! {f(t)\, dt \over t^2\! -\!x^2}dt =\! \int_0^a \!\!
     {f(t)\! -\! f(x) \over t^2-x^2 }\, dt - {f(x) \over 2x}
     \ln {a+x \over a-x } \,.
    \end{eqnarray}
  Then we divide the integration interval into pieces bounded
by the remaining singularities, $0 \le t \le b$, $b \le t \le a$,
and  $a \le t \le 1$. In each interval we substitute the
integration variable by an appropriate function $t=t(u)$ and
integrate over $u$ such that the new integrand has no infinity and
vanishes rapidly at the boundaries. This new integral may thus be
evaluated as a sum over an equidistant grid $u_i$ with constant
weights. For example we write
     \begin{eqnarray}  
     \int_0^{\tau} \!\! g(t)dt = \int_0^1 \!\! g[t(u)]\, t'(u)\, du
     \approx \sum_{i=1}^N g_i w_i
     \end{eqnarray}
with $g_i=g[t(u_i)]$, $u_i = (i-1/2)/N$, $w_i=t'(u_i)/N$,
$t'(u) = dt/du$, and $i=1,\, 2,\, 3,\, \dots\, N$.
This integration method is very accurate if the substitution is
chosen such that the weights $w_i$ and the products $g_iw_i$
vanish rapidly at the integration boundaries, e.g.,
$w_i\sim u_i^p$ and $w_i\sim (1-u_i)^q$ with $p \gg 1$ and $q \gg 1$.
Simple choices of this substitution in the example (A2) are
     \begin{eqnarray}  
     t(u) = (3u^2 - 2u^3) \tau, ~~ t'(u) = 6u(1-u) \tau \,,
     \end{eqnarray}
or better,
     \begin{eqnarray}  
     t(u) = (10u^3\! -\! 15u^4\! +\! 6u^5) \tau \,, ~~
     t'(u) =30u^2(1 \! -\! u)^2 \tau \,.
     \end{eqnarray}
Higher accuracy is achieved by the following substitution. We chose
equidistant $u_i = (i-1/2)/N$ as above and then iterate (A3) $m$
times starting with $s_i = u_i$ and $w_i=\tau/N$ according to
     \begin{eqnarray}  
     w:= 6(s-s^2)w \,, ~~ s:= 3s^2 -2s^3  ~~~~ (m {\rm ~times}) \,.
     \end{eqnarray}
 Finally we write $t(u_i) =s_i \tau $.
 The weights $w_i = t'(u_i)/N$ of this substitution vanish at
 the boundaries with exponents $p=q=2^{(m-1)}$, which can be made
 arbitrarily large. For example, using $m=5$ iterations one gets
 the exponents $p=q=2^4=16$.

   An infinity $g(t) \propto 1/t^\eta$ in the original integral (A2)
 leads, after this substitution, to a new integrand vanishing at
 $t=0$ as $g[t(u)] t'(u) \propto u^\vartheta$ with
 $\vartheta =p(1-\eta)-\eta$.  Thus, for the example
 $\eta = 1/2$ with $p=16$ the new integrand near
 $u=0$ vanishes as $u^{7.5}$ and the terms in the sum (A2) as
 $(i-1/2)^{7.5}$, in spite of the singular original integrand.
 For general exponent $\eta$, to reach high accuracy one should
 choose $m$ so large that the new exponent is
 $\vartheta = (1-\eta)2^{m-1} -\eta \ge 4$, or approximately
 $m \ge 3.5 - 1.5\ln(1-\eta)$. To avoid spurious results due to
 rounding errors, one has to add in all vanishing denominators
 a small $\epsilon \approx 10^{-15}$ by writing, e.g.,
 $(|t^2 -b^2| +\epsilon )^\beta$.

 In the limit of a
 large negative slope $\gamma \to -\infty$ one has $\beta \to 1$ and
 the integrals (17,18) containing a factor $|t^2 - b^2|^{-\beta}$
 are close to diverging. In this case the singular part in these
 integrals should be integrated analytically, similar as shown in
 Eq.~(A1). The subtracted terms are conveniently chosen such that
 the integral which has to be taken analytically is simple, e.g.,
 $\int t\cdot (b^2-t^2)^{-\beta} dt$. Note that the numerator
 $f(t)$ in Eqs.~(15--19,22) is discontinuous at $t=b$.

 \end{multicols}
 \end{document}